\documentclass[aps,pre,twocolumn,twoside,showpacs]{revtex4}
\usepackage{graphicx,amssymb,amsmath,wasysym,color}
\usepackage{multirow}

\newcommand{\dphii}{{\frac{d \phi_i}{dt}}}
\newcommand{\be}{\begin{equation}}
\newcommand{\ee}{\end{equation}}
\newcommand{\bea}{\begin{eqnarray}}
\newcommand{\eea}{\end{eqnarray}}
\newcommand{\bds}{\begin{displaystyle}}
\newcommand{\eds}{\end{displaystyle}}
\newcommand{\balign}{\begin{align}}
\newcommand{\ealign}{\end{align}}
\newcommand{\couplingcoeff}{J}
\newcommand{\orderparameter}{r}
\newcommand{\localfield}{\bar{r}}
\newcommand{\stationary}{\rm (s)}
\newcommand{\sceqs}{(\ref{eqn:localfield-expand})
and (\ref{eqn:orderparameter-expand}) }

\begin{document}
\title{Synchronization transition of heterogeneously coupled
oscillators\\ on scale-free networks}
\author{E.~Oh$^1$, D.-S.~Lee$^2$, B.~Kahng$^{1,3}$ and D.~Kim$^1$\\}
\affiliation{\mbox{$^1$ CTP {\rm \&} FPRD, School of Physics and
Astronomy, Seoul National University, Seoul 151-747, Korea}\\
\mbox{$^2$ Theoretische Physik, Universit\"{a}t des Saarlandes,
  66041 Saarbr\"{u}cken, Germany}\\
\mbox{$^3$ Center for Nonlinear Studies, Los Alamos National
Laboratory, Los Alamos, NM 87545}}
\date{\today}
\begin{abstract}
We investigate the synchronization transition of the modified
Kuramoto model where the oscillators form a scale-free network
with degree exponent $\lambda$. An oscillator of degree $k_i$ is
coupled to its neighboring oscillators with asymmetric and
degree-dependent coupling in the form of $\couplingcoeff
k_i^{\eta-1}$. By invoking the mean-field approach, we determine
the synchronization transition point $J_c$, which is zero (finite)
when $\eta > \lambda-2$ ($\eta < \lambda-2$). We find eight
different synchronization transition behaviors depending on the
values of $\eta$ and $\lambda$, and derive the critical exponents
associated with the order parameter and the finite-size scaling in
each case. The synchronization transition is also studied from the
perspective of cluster formation of synchronized vertices. The
cluster-size distribution and the largest cluster size as a
function of the system size are derived for each case using the
generating function technique. Our analytic results are confirmed
by numerical simulations.
\end{abstract}
 \maketitle

\section{Introduction}
\label{sec:intro} Synchronization of oscillations is one of the
fundamental nonlinear phenomena in biology, physics, chemistry,
communication science, and many other branches of science and
engineering~\cite{book}. Recently, the dynamics of synchronization
of oscillators located at each vertex in complex networks has
attracted much attentions. That is because the small-world feature
of complex networks is closely related to their synchronizability.
By the small-world feature, we mean that the average separation
$\langle d \rangle$ between a pair of vertices scales at most
$\langle d \rangle \sim \log N$, where $N$ is the number of
vertices in the system. It was shown that a small-world network
model introduced by Watts and Strogatz~\cite{watts98} is more
synchronizable than regular lattice~\cite{pecora}. However, such
feature is not observed in scale-free (SF) networks. SF networks
are the networks that exhibit a power-law degree distribution
$P_d(k) \sim k^{-\lambda}$ and degree $k$ is the number of edges
connected to a given vertex~\cite{barabasi99}. In SF networks, the
heterogeneity in the degree distribution suppresses their
synchronizability~\cite{synch_SF1}. Thus, it was desired to
introduce a new dynamic model that prompts SF networks to be more
synchronizable.

The dynamics of synchronization is described by various forms of
coupled equations. A linearly coupled model is probably the
simplest one. In the model, $N$ oscillators are coupled when they
are connected via edges. Coupling constant is normally symmetric;
however, it is not necessarily symmetric to achieve a better
synchronizability. This case can happen in SF networks: It was
shown recently~\cite{motter05} that the synchronizability becomes
maximum when information flow diffuses and reaches a uniform
stationary state over the entire system. Here, the mapping from
synchronization dynamics to information flow can be naturally
introduced, because the linearly coupled equation is nothing but
the diffusion equation. It was shown that the uniform-stationary
state can be reached by introducing asymmetric and weighted
coupling strength between a pair of vertices or oscillators.

To be specific, the dynamic model with the asymmetric coupling
strength is written as
\begin{equation} \frac{d\phi_i}{dt}= f(\phi_i)
-\frac{J}{k_i^{1-\eta}}\sum_{j=1}^{N} a_{ij}\left(
h(\phi_i)-h(\phi_j)\right)\label{eqn:linear}
\end{equation}
for $i=1, \ldots, N$. Here, $\phi_i$ is the phase of an oscillator
located at vertex $i$, $f(\phi)$ describes the dynamics of an
individual oscillator, $J$ is the overall coupling strength. $k_i$
is the degree of vertex $i$, and $a_{ij}$ is an element of the
adjacent matrix, which is 1 if vertices $i$ and $j$ are connected
and 0 otherwise. $h(\phi_i)$ is the output function and take a
form of $h(\phi_i)=\phi_i$ for the linear case. It is noteworthy
that the coupling strength of Eq.~(\ref{eqn:linear}) is asymmetric
and weighted due to the factor $1/k_i^{1-\eta}$ unless $\eta=1$.
When $\eta>0$, vertices with large degree can influence other
vertices significantly on regulating phases due to their large
numbers of connections; on the other hand, when $\eta<0$, the
influence is reduced. It was found~\cite{motter05} that the system
is most synchronizable when $\eta=0$, irrespective of the value of
the degree exponent of a given SF network.

In this paper, we study the pattern of synchronization transition
for a modified Kuramoto model~\cite{kuramoto}, which is similarly
modified with the asymmetric and weighted coupling strength as
\begin{equation}
\dphii = \omega_i - \frac {\couplingcoeff} {k_i^{1-\eta}}
\sum_{j=1}^{N} a_{ij}\sin(\phi_i-\phi_j).\label{eqn:eom}
\end{equation}
The oscillators are located at each vertex $i=1,\dots,N$ of a SF
network with degree exponent $\lambda$. Here, $\omega_i$ is the
natural frequency of the $i$th oscillator selected from the
Gaussian distribution $g(\omega)=e^{-\omega^2/2}/\sqrt{2\pi}$. We
find that the modified Kuramoto dynamic model displays a very
complex and rich behaviors in the space of the two tunable
parameters ($\eta,\lambda$).

The synchronization transition from a desynchronized to a
synchronized state occurs at the critical point $J_c$. For small
$J \ll J_c$, the coupling strength is so weak that an individual
vertex maintains its own phase different from others; therefore,
the entire system is desynchronized. As the coupling strength $J$
increases, a cluster of vertices is more likely to be coupled, to
be in a common or almost the same phase, and thus forms a cluster
of synchrony. Size of such clusters becomes diverse as the
coupling strength $J$ increases. At the critical point $J_c$, the
system reaches a self-organized state, and the cluster-size
distribution follows a power law,
\begin{equation}
n(s)\sim s^{-\tau-1}\label{cluster} \end{equation} in the
thermodynamic limit. For $J \gg J_c$, the power-law behavior no
longer holds and the entire system is synchronized.

The order parameter of the synchronization transition is defined
as
\begin{equation}
\label{eqn:glovalop} \orderparameter e^{i\theta} =
\frac{1}{N}\sum_{i=1}^N e^{i\phi_i}.
\end{equation}
In the synchronized state, the phases $\phi_i$ of each vertex are
narrowly distributed around an average phase $\theta$. The
amplitude $r$ of the order parameter has a finite value; on the
other hand, $r\approx 0$ in the desynchronized state. Thus, the
exponent $\beta$ associated with the order parameter is defined
via the relation, \begin{equation} r\sim \Delta^{\beta},
\end{equation} where $\Delta=(J-J_c)/J_c$. In finite-size systems,
the order parameter is described in terms of a scaling function as
\be r \sim N^{-\beta/\mu}\psi(\Delta N^{1/\mu}).\ee

In the recent works~\cite{synch_SF2,synch_SF5}, the nature of the
transitions and the finite-size scalings have been studied for the
case of $\eta=1$. In this work, we determine the order parameter
and the size distribution of synchronized clusters for general
$\eta$ using the mean-field approach and the generating function
technique. Moreover, we construct a finite-size scaling function
for the order parameter, and determine the exponent $\mu$. Even
for a simple extension of $\eta\ne 1$, we find that the obtained
result is very rich. There exist eight distinct transition
behaviors depending on the values of $\eta$ and $\lambda$.
Therefore, the result can be helpful in understanding diverse
dynamic phenomena arising on SF networks.

The paper is organized as follows: In Sec.~\ref{sec:meanfield}, we
first introduce and apply the mean-field approach to the dynamic
equation (\ref{eqn:eom}). We construct a self-consistent equation
for a local field and determine the order parameter. Next, the
critical point is determined and the behavior of the order
parameter near the critical point is obtained in
Sec.~\ref{sec:anal1}. The size distribution of synchronized
clusters and the largest cluster size at the critical point are
solved in Secs.~\ref{sec:anal2} and \ref{sec:fss}, respectively.
The finite-size scaling analysis for the order parameter is
performed and the results are checked numerically in
Sec.~\ref{sec:nu}. Summary and discussion follow in
Sec.~\ref{sec:conclusion}.

\section{order parameter equation}
\label{sec:meanfield}

In this section, we analyze the modified Kuramoto equation
(\ref{eqn:eom}) in the framework of the mean-field approach by
constructing a self-consistent equation for a local field. To
proceed, we define $\localfield_i$ and $\bar{\theta}_i$ as the
amplitude and phase of the local field at vertex $i$,
respectively, via
\begin{equation} \label{eqn:localop} \localfield_i
e^{i\bar{\theta}_i} = \frac {1} {k_i} \sum_{j=1}^N
a_{ij}e^{i\phi_j}.
\end{equation} Then,
Eq.~(\ref{eqn:eom}) is rewritten in terms of the local field as
\begin{equation}
\label{eqn:localemo} \dphii =
   \omega_i - \couplingcoeff \localfield_i k_i^{\eta} \sin (\phi_i -
       \bar{\theta_i}).
\end{equation}
Once the amplitude $\localfield_i$ and the phase $\bar{\theta}_i$
of the local field are determined, one can solve
Eq.~(\ref{eqn:localemo}) easily. The local field $\localfield_i$
is determined in a self-consistent manner.

We consider the probability density
$\rho_i^{(s)}(\phi|\omega)d\phi$ that the phase of an oscillator
$i$ with natural frequency $\omega$ lies between $\phi$ and
$\phi+d\phi$ in the steady state~\cite{strogatz_physica}. Using a
previous result \cite{strogatz_physica} that
$\rho_i^{(s)}(\phi|\omega)d\phi$ is inversely proportional to the
speed of $\phi$, one can obtain that \be \rho_i^{\stationary}
(\phi|\omega) =
\begin{cases}
\delta \left[ \phi -\bar{\theta}_i-  \sin^{-1} \left( \displaystyle{
    \frac{\omega}{\omega_{*,i}}}\right)\right] &
{\rm if  }\ |\omega|\leq \omega_{*,i}, \\
&\\
\displaystyle{ \frac{\sqrt{\omega^2 - \omega_{*,i}^2}}{ 2\pi | \omega -
\omega_{*,i} \sin (\phi-\bar{\theta}_i) |}} & \rm{otherwise},
\end{cases}
\label{eqn:rho} \ee where $\omega_{*,i} = \couplingcoeff
\localfield_i k_i^\eta$. This result implies that an oscillator
$i$ with natural frequency $\omega$ has its phase that is locked
at $\phi=\bar{\theta}_i+\sin^{-1}(\omega/\omega_{*,i})$ and
$d\phi_i/dt=0$ if $|\omega|\leq \omega_{*,i}$. Otherwise, its
phase drifts with a finite speed, $d\phi_i/dt \ne 0$. Next, we can
evaluate the order parameter using the stationary probability
density in Eq.~(\ref{eqn:rho}) as \be r
e^{i\theta}=\frac{1}{N}\sum_i \int_{-\infty}^{\infty} d\omega
g(\omega) \int d\phi \rho_i^{(s)} (\phi|\omega) e^{i\phi}.
\label{eqn:glovalop-decomp} \ee

Although $\localfield_i$, $\bar{\theta}_i$, and
$\rho_i^{\stationary}(\phi|\omega)$ can fluctuate over $i$ in the
steady state, we assume here that they depend only on degree
$k_i$. This is a mean field approximation. Keeping only the
degree-dependent fluctuations, one can obtain a self-consistent
equation for the local field through Eq.~(\ref{eqn:localop}) as
\begin{align}
\localfield(k) e^{i\bar{\theta}(k)} &= \sum_{k'=1}^{k_m} P(k'|k)\nonumber\\
&~~~\times
\int_{-\infty}^{\infty} d\omega g(\omega)
       \int_0^{2\pi} d\phi \rho^{\stationary}(\phi|\omega,k') e^{i\phi},
\label{localfield}
\end{align}
where $\rho^{\stationary}(\phi| \omega,k)$ is given by the right
hand side of Eq.~(\ref{eqn:rho}) with
$\omega_*(k)=\couplingcoeff\localfield(k) k^\eta$ replacing
$\omega_{*,i}$. $P(k^{\prime}|k)$ denotes the probability that a
neighboring vertex of a given vertex with degree $k$ has degree
$k^{\prime}$, and $k_m$ is the natural cutoff of degree. Here, we
consider only the case that the network is random and does not
have any type of degree-degree correlation; then,
$P(k^{\prime}|k)$ can be written as
$k^{\prime}P_d(k^{\prime})/\langle k\rangle$ with $\langle
k\rangle = \sum_k k P_d(k)$. After that, one can see that both
$\localfield(k)$ and $\bar{\theta}(k)$ are independent of degree
$k$, and therefore, we can drop the $k$-dependence in
$\localfield$ and $\bar{\theta}$ from now on.

The last integral of Eq.~(\ref{localfield}) is evaluated as
\begin{align}
&\int_{0}^{2\pi}  d\phi \rho^{(s)}(\phi|\omega,k) e^{i\phi}\nonumber \\
&= e^{i\bar{\theta}}
\begin{cases}
i(\omega/\omega_*(k))-i\sqrt{(\omega/\omega_*(k))^2-1}, &(\omega > \omega_*(k)), \\
i(\omega/\omega_*(k)) + \sqrt{ 1-(\omega/\omega_*(k))^2},  &(|\omega|\leq  \omega_*(k)), \\
i(\omega/\omega_*(k)) + i \sqrt{ (\omega/\omega_*(k))^2 - 1},
&(\omega < -\omega_*(k)).
\end{cases}
\end{align}
The remaining integration in Eq.(\ref{localfield}) for $\omega
> \omega_*(k)$ and $\omega < \omega_*(k)$ cancel out due to
the fact $g(\omega)=g(-\omega)$. As a result, only the oscillators
having the frequency within the range $|\omega|\leq \omega_*(k)$
contribute to the local field in Eq.~(\ref{eqn:localop}). Thus,
one obtains \be \label{eqn:localopcontinum} \localfield =
\sum_{k=1}^{k_m} \frac{k P_d(k)}{ \langle k \rangle}
\int_{-\omega_*(k)}^{\omega_*(k)} d\omega g(\omega)
  \sqrt{1-
  \left(\displaystyle{\frac{\omega}{\omega_*(k)}}\right)^2},
\ee which is the self-consistent equation for $\localfield$. Note
that $\localfield$ is contained in $\omega_*(k)=J{\bar
r}k^{\eta}$. After the local field is obtained, the order
parameter in Eqs.~(\ref{eqn:glovalop}) or
(\ref{eqn:glovalop-decomp}) is calculated as \be
\label{eqn:glovalopcontinuum} \orderparameter=
\sum_{k=1}^{k_m}P_d(k)\int_{-\omega_*(k)}^{\omega_*(k)} d\omega
g(\omega)
 \sqrt{1-\left(\displaystyle{\frac{\omega}{\omega_*(k)}}\right)^2}.
\ee

\section{Synchronization transition}
\label{sec:anal1} In this section, we solve the self-consistent
equation, Eq.~(\ref{eqn:localopcontinum}) explicitly, and then
investigate the behavior of the order parameter near the critical
point via Eq.~(\ref{eqn:glovalopcontinuum}). To proceed, we first
recall that the degree distribution is given in a closed form as
$P_d(k)=k^{-\lambda}/H_{k_m}^{\lambda}$ for $\lambda>2$, where
$H_{k_m}^\lambda$ is the generalized harmonic number, defined by
$H_m^q \equiv \sum_{k=1}^{m} k^{-q}$ and $k_m \sim
N^{1/(\lambda-1)}$. Substituting
$g(\omega)=e^{-\omega^2/2}/\sqrt{2\pi}$ to
Eq.~(\ref{eqn:localopcontinum}), one can derive the local field
$\localfield$ as
\begin{align}
\label{eqn:localfield-expand} \localfield & = \sum_{k=1}^{k_m}
\frac {k P_d(k)} {\langle k \rangle} \omega_*(k)
\int_{-\pi/2}^{\pi/2} d\phi \cos^2\phi \frac {1}{\sqrt{2 \pi}}
e^{-{[\omega_*(k) \sin\phi]^2}/{2}}
\nonumber\\
&=  \sum_{n=0}^{\infty}\frac{ (n-1/2)!\,(-1)^n \,
H_{k_m}^{\lambda-\eta-2n \eta-1}}{n!\,(n+1)!\,2^{n+3/2}\,
H_{k_m}^{\lambda-1}} (\couplingcoeff\localfield)^{2n+1}
 \nonumber\\
&\equiv \sum_{n=0}^\infty \bar{A}_n (\couplingcoeff \localfield)^{2n+1},
\end{align}
where we used the Taylor expansion of
$e^{-(\omega_*(k)\sin\phi)^2/2}$ and the integration
$\int_{-\pi/2}^{\pi/2} d\phi \cos^2\phi \sin^{2n}\phi = \pi^{1/2}
(n-1/2)! /\left(2(n+1)!\right)$. Similarly, the order parameter is
evaluated as
\begin{align}
\orderparameter &= \sum_{n=0}^{\infty} \frac{ (n-1/2)! \,\,(-1)^n
H_{k_m}^{\lambda-\eta-2n \eta}} { 2^{n+3/2}\,\, n!\,(n+1)!\,\,
H_{k_m}^\lambda}
(\couplingcoeff\localfield)^{2n+1}\nonumber\\
&\equiv \sum_{n=0}^\infty A_n (\couplingcoeff \localfield)^{2n+1}.
\label{eqn:orderparameter-expand}
\end{align}

If $\eta\leq 0$, the generalized harmonic numbers in $\bar{A}_n$
and $A_n$ are finite, and then they can be represented in terms of
the Riemann zeta functions for all $n$ below and they are denoted
as $\bar{B}_n$ and $B_n$, respectively. That is,
\begin{equation}
\bar{A}_n\approx \bar{B}_n = \frac{ (n-1/2)! (-1)^n
\zeta(\lambda-\eta-2n \eta-1)} { 2^{n+3/2} n!(n+1)!
\zeta(\lambda-1)} \label{eqn:BBbar}
\end{equation}
and
\begin{equation}
{A}_n\approx {B}_n = \frac{ (n-1/2)! (-1)^n \zeta(\lambda-\eta-2n
\eta)} { 2^{n+3/2} n!(n+1)! \zeta(\lambda)}, \label{eqn:BB}
\end{equation}
respectively. Using these formulae, the local field and the order
parameter are determined by Eqs.~\sceqs.

On the other hand, it is remarkable that when $0 < q <1$, the
generalized harmonic number diverges as $H_m^q \simeq
(m+1)^{1-q}/(1-q)+\zeta(q)+{\cal O}(m^{-q})$ in the $m\to\infty$
limit, which is shown in Appendix. Then, Eqs.~\sceqs are divided
into analytic and singular parts as
\begin{equation}
\localfield=\sum_n \bar{B}_n (\couplingcoeff
\localfield)^{2n+1}+ \bar{C}(\couplingcoeff \localfield k_m^\eta)
  (\couplingcoeff \localfield)^{(\lambda-2)/\eta}
  \label{eqn:localrs}
\end{equation} and
\begin{equation}
\orderparameter=\sum_n B_n (\couplingcoeff \localfield)^{2n+1}+
{C}(\couplingcoeff \localfield k_m^\eta)
  (\couplingcoeff \localfield)^{(\lambda-1)/\eta},
  \label{eqn:globalrs}
\end{equation}respectively, where the functions $\bar{C}(x)$ and
${C}(x)$ are defined in Appendix. In the $x\to\infty$ limit
corresponding to the thermodynamic limit, $\bar{C}(x)$ and $C(x)$
reduce to $\bar{C}_\infty$ and $C_\infty$, respectively, defined
as
\begin{align}
\bar{C}_\infty&=\displaystyle{
\frac{[(\lambda-2\eta-2)/2\eta]![(2-\lambda-\eta)/2\eta]!} {\eta
2^{(\lambda+4\eta-2)/2\eta}[(\lambda+\eta-2)/2\eta]! \zeta(\lambda-1)}},
\nonumber\\
{C}_\infty&=\displaystyle{
\frac{[(\lambda-2\eta-1)/2\eta]![(1-\lambda-\eta)/2\eta]!} {\eta
2^{(\lambda+4\eta-1)/2\eta}[(\lambda+\eta-1)/2\eta]!
\zeta(\lambda)}}. \label{eqn:c0}
\end{align}
Thus, the local field and the order parameter are written as
 \be \localfield = \sum_{n=0}^{\infty}
\bar{B}_n(\couplingcoeff\localfield)^{2n+1}
   + \bar{C}_\infty(\couplingcoeff\localfield)^{(\lambda-2)/\eta} +\ldots,
\label{eqn:local-pos-eta}
\ee
and
\begin{align}
\label{eqn:orderparameter-pos-eta}
\orderparameter=  \sum_{n=0}^{\infty} B_n (\couplingcoeff
\localfield)^{2n+1} + C_\infty (\couplingcoeff \localfield)^{(\lambda-1)/\eta} + \ldots,
\end{align}
for $\couplingcoeff\localfield k_m^\eta\gg 1$. We remark that the
singular terms appear only in the limit $\couplingcoeff
\localfield k_m^\eta\to \infty$. For the case of
$\couplingcoeff\localfield k_m^\eta\ll 1$, however, Eqs.~\sceqs
are valid.

Next, we determine the critical point. To proceed, we investigate
the behavior of the local field as a function of $\couplingcoeff$,
which depends on the sign of $\eta$.

{\bf (i) In the case of $\eta\leq 0$,} $\bar{A}_n$ and $A_n$ are
finite. One can see from Eq.~(\ref{eqn:localfield-expand}) that
the local field is zero for $\bar{A}_0 J< 1$ and  non-zero for
$\bar{A}_0 J> 1$. The order parameter behaves in the same manner
as that of the local field from
Eq.~(\ref{eqn:orderparameter-expand}). Thus, we obtain the
critical point as \be \couplingcoeff_c = \frac{1}{\bar{A}_0} =
\frac {2\sqrt{2}}{\sqrt{\pi}}
      \frac {H_{k_m}^{\lambda-1}}{H_{k_m}^{\lambda-1-\eta}}.
\label{eqn:Jc} \ee As $\lambda\to\infty$, the critical point $J_c$
approaches $2\sqrt{2}/\sqrt{\pi} \simeq 1.60$ in the limit
$N\to\infty$, which is consistent with that found in case of the
globally-coupled oscillators~\cite{kuramoto}.

When $\couplingcoeff > \couplingcoeff_c$, the local field
$\localfield$ and the order parameter $\orderparameter$ are
non-zero. When $\couplingcoeff$ is close to $\couplingcoeff_c$,
\be \localfield \simeq (|\bar{A}_1| \couplingcoeff_c^3)^{-1/2}
\Delta^{1/2}\label{r10}\ee and \be\orderparameter\simeq {A}_0
(|\bar{A}_1| \couplingcoeff_c)^{-1/2} \Delta^{1/2},\ee where
$\Delta= {(\couplingcoeff-\couplingcoeff_c)}/{\couplingcoeff_c}$.
Thus, we obtain that $\beta=1/2$. Again, this result is consistent
with the one obtained from the globally-coupled
oscillators~\cite{kuramoto}.

{\bf (ii) In the case of $\eta>0$,} the singular terms in
Eqs.~(\ref{eqn:local-pos-eta}) and
(\ref{eqn:orderparameter-pos-eta}) can be crucial in determining
the critical point and the order parameter. Depending on relative
magnitude of $\lambda$ and $\eta$, we divide the case of $\eta
>0$ into four subcases:
\begin{itemize}
\item[(I)] When $0<\eta<(\lambda-2)/3$ (i.e., $\lambda > 3\eta+2$),
$\localfield \simeq \bar{B}_0 \couplingcoeff \localfield +
\bar{B}_1 \couplingcoeff^3 \localfield^3 + \cdots$ for small
$\localfield$ from Eq.~(\ref{eqn:local-pos-eta}). Then
$\couplingcoeff_c$ and $\localfield$ behave as those for $\eta <
0$ presented in Eqs.~(\ref{eqn:Jc}) and (\ref{r10}). \item[(II)]
When $(\lambda-2)/3 <\eta<\lambda-2$ (i.e., $\eta+2 < \lambda <
3\eta+2$), the dominant contribution is made from the singular
term of Eq.~(\ref{eqn:local-pos-eta}). Then \be \localfield \simeq
\bar{B}_0 \couplingcoeff\localfield+\bar{C}_\infty
(\couplingcoeff\localfield)^{(\lambda-2)/\eta}+\ldots, \ee leading
to \be J_c\sim
1/\bar{B}_0=\frac{2\sqrt{2}}{\sqrt{\pi}}\frac{\zeta(\lambda-1)}{\zeta(\lambda-\eta-1)}
\ee and  \be \localfield\sim \orderparameter \sim
\Delta^{\eta/(\lambda-2-\eta)}. \ee
\item[(III)] When $\lambda-2 < \eta < \lambda-1$ (i.e., $\eta+1
< \lambda < \eta+2$), the critical point in Eq.~(\ref{eqn:Jc}) for
finite $N$ behaves as \be J_c \sim k_m^{-(\eta-\lambda+2)}\sim
N^{-(\eta-\lambda+2)/(\lambda-1)}.\ee Thus, it approaches zero in
the thermodynamic limit. $\localfield$ is always positive unless
$\couplingcoeff$ is zero as $\localfield\sim
\couplingcoeff^{(\lambda-2)/(\eta-\lambda+2)}$ for small $J$ and
$\orderparameter \sim J\localfield \sim
\couplingcoeff^{\eta/(\eta-\lambda+2)}$.

\item[(IV)] When $\eta > \lambda-1$ (i.e., $\lambda < \eta+1$),
we obtain that $\orderparameter \sim (\couplingcoeff
\localfield)^{(\lambda-1)/\eta}$. Using the result of
$\localfield$ obtained in (III), we obtain that \be
\orderparameter \sim
\couplingcoeff^{(\lambda-1)/(\eta-\lambda+2)}. \ee
\end{itemize}

We summarize the result as follows: When $\eta < \lambda-2$ (in
the (I) and (II) cases), the critical point $J_c$ is finite;
however, when $\eta > \lambda-2$ (in the (III) and (IV) cases),
$J_c=0$ in the thermodynamic limit $N\to \infty$. Thus, the
critical exponent $\beta$ associated with the order parameter is
defined through the relation, $r\sim \Delta^{\beta}$ ($r\sim
J^{\beta}$) for the former (latter) case. The exponent $\beta$ is
evaluated in each case as follows: \be \beta =
\begin{cases}
1/2 & \text{in~(I),}\\
\eta/(\lambda-2-\eta) & \text{in~(II),}\\
\eta/(\eta-\lambda+2) & \text{in~(III),}\\
(\lambda-1)/(\eta-\lambda+2) & \text{in~(IV).}
\end{cases}
\label{eqn:beta}
\ee

The result of the critical point is consistent with those of other
phase transition problems such as the percolation transition and
the epidemic spreading in SF networks. Moreover, the result for
the case of $\eta=1$ reduces to the previous result
\cite{synch_SF2,synch_SF5}. Moreover, the result $\beta=1/2$ for
$\eta=1$ and $\lambda > 5$ is reduced to the mean-field result in
regular lattice.

\section{Cluster formation of synchronized oscillators}
\label{sec:anal2}

In this section, we investigate in detail how the coupled
oscillator system develops its synchrony as the coupling strength
increases. To this end, we study the formation of clusters
comprising synchronized vertices as a function of the coupling
strength $J$. We use the generating function approach to derive
the cluster-size distribution.

\subsection{Cooperative versus background synchrony}
\label{sec:synchrony}

The order parameter averaged over the natural frequency
distribution $g(\omega)$ can be written  as \be
\orderparameter=\frac{1}{N}\sqrt{\sum_{i}\langle \cos^2\phi_i
+\sin^2\phi_i \rangle+\sum_i \sum_{j \ne i}\langle
\cos(\phi_i-\phi_j)\rangle}\label{eqn:order-anal-J0} \ee from
Eq.(\ref{eqn:glovalop}). Here, the brackets represent the average
over $g(\omega)$. For the case of $\couplingcoeff=0$, each element
oscillates independently, so that $\langle
\cos(\phi_i-\phi_j)\rangle=0$ for $i \ne j$. Thus, the order
parameter is evaluated as \be \orderparameter_{\couplingcoeff=0}
\sim \frac{1}{\sqrt{N}}. \label{eqn:order-anal-J0} \ee As
$\couplingcoeff$ increases, clusters comprising synchronized
oscillators are more likely to form. We here define a cluster as
a group of vertices (or oscillators) which are connected and in
the same coherent state: Two oscillators are coherent if its
time-average correlation function $C_{ij}$, defined as \be
C_{ij}=\frac{1}{(t_1-t_0)} \sum_{t=t_0+1}^{t_1} \langle \cos
(\phi_i(t) - \phi_j(t)) \rangle,\ee is larger than a preassigned
threshold value $C_{\rm th}$. We choose $C_{\rm th}$ value to
generate the cluster-size distribution in a power law form at the
critical point $J_c$. As such clusters form, the term of $\sum_i
\sum_{j \ne i}\langle \cos(\phi_i-\phi_j)\rangle$ becomes nonzero
and dominant. The order parameter is then evaluated as \be
\orderparameter \sim \frac{\sqrt{\sum_{\kappa} s^2_{\kappa}}}{N},
\label{eqn:order-anal-J-pos} \ee where $\kappa$ is the index of
cluster and $s_{\kappa}$ is the size of cluster $\kappa$, i.e.,
the number of vertices within the cluster $\kappa$. Note that
$\sum_{\kappa} s_{\kappa}=N$, and Eq.~(\ref{eqn:order-anal-J-pos})
reduces to Eq.~(\ref{eqn:order-anal-J0}) when $\couplingcoeff=0$
because each cluster size is 1. In the synchronized state when
$\couplingcoeff \gg \couplingcoeff_c$, the size of the largest
cluster, denoted as $S$, is of ${\cal O}(N)$, and thus the order
parameter is approximately given as \be \orderparameter \sim S/N.
\ee Next, we study the cluster-size distribution and the size of
the largest cluster as a function of $J$.

\begin{figure*}
\resizebox*{!}{5cm}{\includegraphics{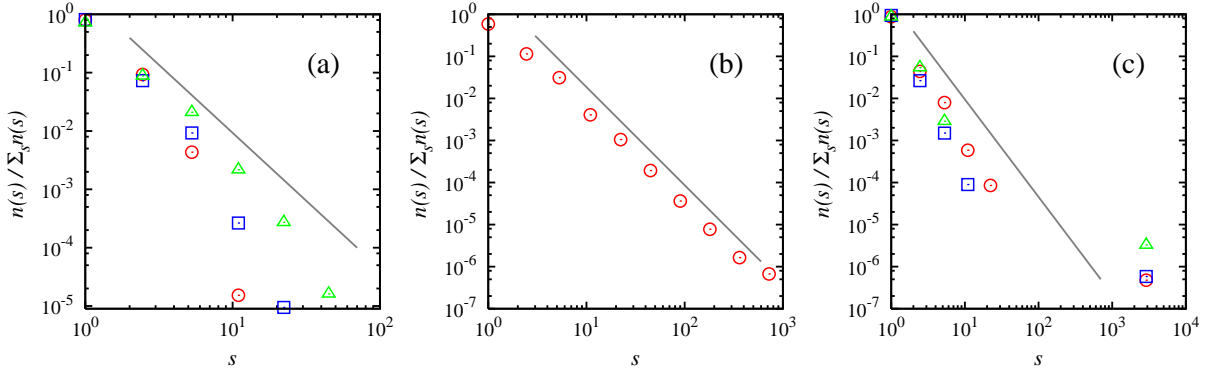}} \caption { (Color
online) Cluster-size distributions $n(s)/\sum_s n(s)$ of
synchronized clusters for the networks generated with
$\lambda=4.0$, $\eta=0.0$, and $N=3000$ at $ J=0.7J_c $
(\textcolor{red}{$\bigcirc$}), $J=0.8J_c$
(\textcolor{blue}{$\square)$}, $J=0.9J_c$
(\textcolor{green}{$\triangle$}) for (a), $J=J_c$
(\textcolor{red}{$\bigcirc$}) for (b), $J=1.1J_c$
(\textcolor{red}{$\bigcirc$}), $J=1.2J_c$
(\textcolor{blue}{$\square$}), and $J=1.3J_c$
(\textcolor{green}{$\triangle$}) for (c), respectively. Solid
lines drawn for reference have a slope of $7/3$ for all. }
\label{fig:cluster}
\end{figure*}

The dynamics of cluster merging with increasing $\couplingcoeff$
results in the change of the cluster-size distribution. Let $n(s)$
be the number of $s$-size clusters. Then $\sum_s sn(s)=N$. The
cluster-size distribution is defined as $n(s)/\sum_s n(s)$. For $J
< J_c$, the cluster-size distribution decays exponentially for
large $s$. However, it decays in the power law form
(\ref{cluster}) at the critical point $J=J_c$, and the associated
exponent $\tau$ depends on the parameters $\eta$ and $\lambda$. We
determine $\tau$ using the generating function method in the next
subsection. For $J
> J_c$, a giant cluster forms and the distribution of finite-size
clusters decays exponentially. The cluster-size distributions for
various values of $J$ are shown in Fig.~\ref{fig:cluster}.

\subsection{Generating function of the cluster-size distribution}

The probability that a vertex belongs to a cluster with size $s$ is
given by $sn(s)/N$, which is denoted as $p(s)$. Invoking the
percolation theory, $p(s)$ follows a power law with an exponential
cutoff, \be p(s)\sim s^{-\tau}e^{-s/s_c}, \label{ps} \ee where $s_c$
is the characteristic size, which depends on $J$ and system size
$N$. In the thermodynamic limit $N\to \infty$, $s_c$ diverges at
$J=J_c$. As in the percolation theory, the generating function
${\cal P}(z)\equiv\sum_s p(s)z^s$ is useful for studying structural
feature of the synchronized clusters, since its singular behavior is
related to the critical behavior of the synchronization transition.
(i) The order parameter $\orderparameter\sim S/N$ can be obtained
from the relation $\orderparameter \simeq \lim_{N\to\infty} [1-{\cal
P}(z_N^*)]$, where ${\cal P}(z_N^*) =\sum_{s<S} p(s)$, i.e., the
contribution by finite-size clusters. This can be achieved by
choosing $z_N^* \approx e^{-1/S_m}$, where $S_m$ is a cluster size
smaller than the largest cluster but larger the second largest
cluster. (ii) From Eq.~(\ref{ps}), one can find that ${\cal P}(z)$
diverges for $z > z_c=\lim_{s\to \infty} p(s)^{-1/s}$, i.e., $z_c
\approx e^{1/s_c}$. Thus, at $J=J_c$, ${\cal P}(z)\sim
(1-z)^{\tau-1}$ as $z\to z_c=1$ in the thermodynamic limit. Thus,
finding the singularity of ${\cal P}(z)$ enables one to obtain
$p(s)$.

For the purpose, we introduce another generating function
$\bar{\cal P}(z)$ as a partner of the local field ${\bar r}$. From
$\bar{\cal P}(z)$, one can define a probability ${\bar p}(s)$ via
the relation $\bar{\cal P}(z)\equiv \sum_s \bar{p}(s)z^s$, where
${\bar p}(s)$ is defined similarly to $p(s)$ as the probability
that a vertex belongs to a synchronized cluster of size $s$
composed of the vertex and $s-1$ neighboring vertices. For finite
$N$, the generating function $\bar{\cal P}(z)$ is analytic for
$|z|\leq 1$ and so is its inverse function $\bar{\cal P}^{-1}(z)$.
To investigate the singularity of $\bar{\cal P}(z)$ near $z=1$, we
consider the expansion of the inverse function $z=\bar{\cal
P}^{-1}(\omega)=1-\sum_{n\geq 1} b_n (1-\omega)^n$ around
$\omega=1$. The coefficient $b_n$ depends on $\couplingcoeff$.
Using Eqs.~(\ref{eqn:localfield-expand}) and
(\ref{eqn:local-pos-eta}) and replacing $\localfield$ by
$1-\omega$, we can find that the generating function $\bar{\cal
P}(z)$ satisfies the self-consistent relations given below:
\begin{align}
z = &\bar{\cal P}(z)  +
\sum_{n=0}^{\infty} \bar{B}_n[\couplingcoeff(1-\bar{\cal P}(z))]^{2n+1}\nonumber\\
  & + \bar{C}_\infty[\couplingcoeff(1-\bar{\cal P}(z))]^{(\lambda-2)/\eta} +\ldots,
\label{eqn:pbar-small-s}
\end{align}
for $\couplingcoeff (1-\bar{\cal P}(z)) k_m^\eta \gg 1$, and \be z
=\bar{\cal P}(z)+\sum_{n=0}^{\infty}
\bar{A}_n[\couplingcoeff(1-\bar{\cal P}(z))]^{2n+1},
\label{eqn:pbar-large-s}\ee for $\couplingcoeff (1-\bar{\cal
P}(z)) k_m^\eta \ll 1$. Similarly, ${\cal P}(z)$ is determined as
\begin{eqnarray}
z-{\cal P}(z)&=&\sum_{n=0}^{\infty} B_n[\couplingcoeff(1-\bar{\cal
P}(z))]^{2n+1} \nonumber
\\&+&C_\infty[\couplingcoeff(1-\bar{\cal
P}(z))]^{(\lambda-1)/\eta}+\ldots, \label{eqn:pandpbar-small-s}
\end{eqnarray}
for $\couplingcoeff (1-\bar{\cal P}(z)) k_m^\eta \gg 1$, and \be
z-{\cal P}(z)=\sum_{n=0}^{\infty} A_n[\couplingcoeff(1-\bar{\cal
P}(z))]^{2n+1}, \label{eqn:pandpbar-large-s} \ee for
$\couplingcoeff (1-\bar{\cal P}(z)) k_m^\eta \ll 1$.

\subsection{Behavior of $p(s)$ at the critical point}

Here, we calculate the probability $p(s)$ to find a vertex in the
$s$-size cluster at the critical
point $J=J_c$ explicitly in each case defined in Sec.\ref{sec:anal1}.\\

{\bf In the case (I),} since $\eta < (\lambda-2)/3$, we obtain
that $z=1+\bar{B}_1 \left(\couplingcoeff_c (1-\bar{\cal
P}(z))\right)^3 + \ldots$ to the leading orders by expanding
$z=\bar{\cal P}^{-1}(\omega)$ around $\omega=1$ in either
Eqs.~(\ref{eqn:pbar-small-s}) or (\ref{eqn:pbar-large-s}). Thus,
we obtain that $1-\bar{\cal P}(z)\sim (1-z)^{1/3}$, leading to \be
\bar{p}(s)\sim s^{-4/3} \ee for large $s$.

Using the obtained leading behaviors of $\bar{\cal P}(z)$ around
$z=1$ in Eqs.~(\ref{eqn:pandpbar-small-s}) and
(\ref{eqn:pandpbar-large-s}), one obtains the behavior of ${\cal
P}(z)$ around $z=1$ as $1-{\cal P}(z) \sim (1-z)^{1/3}$. Thus, \be
p(s)\sim s^{-4/3}. \label{eqn:ps1} \ee

{\bf In the case (II),} the singular term in
Eq.~(\ref{eqn:pbar-small-s}) is relevant. In this case, the
behavior of $\bar{\cal P}(z)$ for $z < z_c$ differs from that for
$z > z_c$. $z_c$ is determined by the criterion $\couplingcoeff
(1-\bar{\cal P}(z_c))k_m^\eta \sim 1$. This case also happens for
the cases (III) and (IV).

In this case, the singular term
$\bar{C}_\infty[\couplingcoeff_c(1-\bar{\cal
P}(z)))]^{(\lambda-2)/\eta}$ is dominant in
Eq.~(\ref{eqn:pbar-small-s}); therefore, it follows that
$1-\bar{\cal P}(z)\sim J_c^{-1}(1-z)^{\eta/(\lambda-2)}$ at
$J=J_c$, which is valid for $z\gg z_c$. From this result,
$\bar{p}(s)$ is obtained as \be \bar{p}(s)\sim
J_c^{-1}s^{-(\eta+\lambda-2)/(\lambda-2)},\label{eqn:pbar<} \ee
which is valid for $s\ll s_c$.

On the other hand, when $J(1-\bar{P}(z))k_m^\eta \ll 1$ so that
$|\bar{A}_{n+1}/\bar{A}_n| [J(1-\bar{P}(z))]^2\ll 1$, one can
obtain that $1-\bar{\cal P}(z)\sim \couplingcoeff_c^{-1}
|\bar{A}_1|^{-1/3}(1-z)^{1/3}$ for $z \ll z_c$; therefore, \be
\bar{p}(s)\sim \couplingcoeff_c^{-1} k_m^{(\lambda-2)/3-\eta}
s^{-4/3} \label{eqn:pbar>} \ee for large $s \gg s_c$. $s_c$ is
evaluated as follows: Substituting the result of $1-\bar{\cal
P}(z)$ in the criterion $\couplingcoeff_c (1-\bar{\cal P}(z_c))
k_m^\eta \sim 1$ and using $(1-z_c) \sim s_c^{-1}$, one can obtain
system-size dependence of the characteristic size $s_c$ explicitly
as \be s_c\sim k_m^{\lambda-2}\sim N^{(\lambda-2)/(\lambda-1)},
\label{eqn:s_c}\ee which diverges as $N\to \infty$.

Together with Eqs.~(\ref{eqn:pbar<}) and (\ref{eqn:pbar>}), we
obtain that \be \bar{p}(s)\sim
\begin{cases}
s^{-(\eta+\lambda-2)/(\lambda-2)} &  (s\ll s_c),\\
k_m^{(\lambda-2)/3-\eta} s^{-4/3} &  (s\gg s_c).
\end{cases}
\ee

Next, using the result of $1-{\cal P}(z)\sim 1-\bar{\cal P}(z)$
obtained from both Eqs.~(\ref{eqn:pandpbar-small-s}) and
(\ref{eqn:pandpbar-large-s}), one can find that $p(s)$ behaves
similarly to $\bar{p}(s)$. That is, \be p(s)\sim
\begin{cases}
s^{-(\eta+\lambda-2)/(\lambda-2)} & (s\ll s_c),\\
k_m^{(\lambda-2)/3-\eta} s^{-4/3} & (s\gg s_c).
\end{cases}
\label{eqn:ps2} \ee

{\bf In the case (III),} the critical point $J_c$ is finite in
finite-size systems as being of order $J_c \sim
k_m^{\lambda-2-\eta}\sim N^{(\lambda-2-\eta)/(\lambda-1)}$.
Plugging the $N$-dependence into Eq.~(\ref{eqn:pbar>}) and the
expression, $1-\bar{\cal{P}}(z) \simeq 1-z +\bar{C}_\infty
[J_c(1-z)]^{(\lambda-2)/\eta}$ for $s\ll s_c$ from
Eq.~(\ref{eqn:pbar-small-s}), one obtains $\bar{p}(s)$ as follows:
\be \bar{p}(s)\sim
\begin{cases}
k_m^{(\lambda-2-\eta)(\lambda-2)/\eta}
s^{-(\lambda-2+\eta)/\eta}& (s\ll s_c),\\
k_m^{-2(\lambda-2)/3} s^{-4/3} & (s\gg s_c).
\end{cases}
\ee

Next, we derive $p(s)$. We find that the leading singular term in
${\cal P}(z)$ for the case $1-z\gg s_c^{-1}\sim k_m^{2-\lambda}$
shows up in two ways. Substituting $1-\bar{\cal P}(z)\approx 1-z +
\bar{C}_\infty [\couplingcoeff_c (1-z)]^{(\lambda-2)/\eta}$ to
Eq.~(\ref{eqn:pandpbar-small-s}), we obtain that $1-{\cal
P}(z)\approx B_0 \couplingcoeff_c (1-z)+B_0 \couplingcoeff_c
\bar{C}_\infty [\couplingcoeff_c
(1-z)]^{(\lambda-2)/\eta}+C_\infty [\couplingcoeff_c
(1-z)]^{(\lambda-1)/\eta}+\cdots$. We compare the second with the
third terms in order of magnitude. Using the fact that $J_c \sim
k_m^{\lambda-2-\eta}$, we find that there exist two subcases for
$s \ll s_c$. The second term $B_0 \couplingcoeff_c \bar{C}_\infty
[\couplingcoeff_c (1-z)]^{(\lambda-2)/\eta}$ is more dominant than
the third term $C_\infty [\couplingcoeff_c
(1-z)]^{(\lambda-1)/\eta}$ when $1-z\ll s_*^{-1}$ and vice versa.
Here, it is found that a new crossover size $s_*$ scales as \be
s_*\sim k_m^{(\eta-1)(\eta-\lambda+2)}. \ee From the behaviors of
${\cal P}(z)$ in the three different subcases, we obtain the
probability $p(s)$ as \be p(s)\sim
\begin{cases}
k_m^{(\lambda-1)(\lambda-2-\eta)/\eta}
s^{-(\lambda-1+\eta)/\eta}& (s\ll s_*),\\
k_m^{(\lambda-2-\eta)(\lambda-2+\eta)/\eta}
s^{-(\lambda-2+\eta)/\eta}& (s_*\ll s\ll s_c),\\
k_m^{(\lambda-2)/3 - \eta} s^{-4/3} & (s\gg s_c).
\end{cases}
\label{eqn:ps3} \ee One can notice that the subcase $s\ll s_*$
diminishes when $\eta\leq 1$, but it is extended as the parameter
$\eta$ increases.

{\bf In the case (IV),} the third term $C_\infty [\couplingcoeff_c
(1-z)]^{(\lambda-1)/\eta}$ in $1-{\cal P}(z)$ in the case (III) is
always dominant when $1-z \gg s_c^{-1}$. Moreover, $A_0$ in
Eq.~(\ref{eqn:pandpbar-large-s}) diverges as $A_0\sim
k_m^{\eta-\lambda+1}$, which has to be considered in the relation,
$1-\bar{P}(z)\simeq A_i J_c (1-\bar{P}(z))$ for $1-z\ll s_c^{-1}$.
Consequently, $p(s)$ behaves as \be p(s)\sim
\begin{cases}
k_m^{(\lambda-1)(\lambda-2-\eta)/\eta}
s^{-(\lambda-1+\eta)/\eta}& (s\ll s_c),\\
k_m^{(1-2\lambda)/3} s^{-4/3} & (s\gg s_c).
\end{cases}
\label{eqn:ps4}
\ee

To substantiate the predictions of this section, we investigate
the asymptotic behavior of $p(s)$ in a numerical manner. The
static model introduced in~\cite{staticmodel} is used for
underlying network in our simulations. The network has $N=3000$
oscillators and its mean degree $\langle k \rangle$ is 4.0. The
values of $\lambda$ and $\eta$ are chosen as 4.0 and 0.0,
respectively. This pair belongs to the case (I). First, we
simulate the system at $J=J_c$ to determine $C_{\rm th}$ defined
in Sec.~\ref{sec:synchrony}. During the simulation, we assume a
large value of $C_{\rm th}$ and then collect the pairs of vertices
that the $C_{ij}$ of each pair is larger than the assumed $C_{\rm
th}$. After that, we determine clusters and obtain the
cluster-size distribution. We then adjust $C_{\rm th}$ by somewhat
decreasing or increasing it, and repeat these procedures until the
power-law distribution appears in the cluster-size distribution.
If the cluster-size distribution follows the power-law form of
$p(s) \sim s^{-\tau}$, the corresponding value of $C_{\rm th}$ is
considered as the threshold value $C_{\rm th}$. It is found
numerically that $C_{\rm th}\approx 0.7$, independent of the
system size $N$. In our simulations, we obtain $\tau+1 \approx
7/3$, which is close to the theoretical value in
Eq.~(\ref{eqn:ps1}), as shown in Fig.~\ref{fig:cluster}(b). We
also performed simulations for various $J < J_c$ and $> J_c$ as
shown in Figs.~\ref{fig:cluster}(a) and (c), respectively. The
power-law behaviors in the cluster-size distribution do not appear
in these cases.

\section{Largest cluster size and finite-size scaling}
\label{sec:fss}

\begin{figure}
\includegraphics[width=\columnwidth]{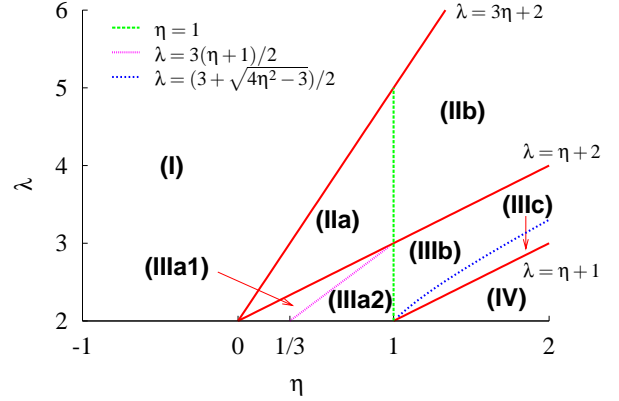}
\caption{(Color online) Diagram in the space of $(\eta,\lambda)$
of eight different domains each corresponding to a distinct
synchronization transition. The transition nature of each domain
is listed in Table~\ref{table:exponents}.} \label{fig:regions}
\end{figure}

\begin{table*}
\caption{The probability to find a vertex in $s$-size cluster
$p(s)$, the largest cluster size $S$ at the critical point, and the
critical exponents $\beta$ and $\mu$ for the eight cases shown in
Fig.~\ref{fig:regions}.} \label{table:exponents}
\begin{tabular}{c|c|c|c|c}
\hline
domain & $p(s)$ & $S$ & $\beta$ & $\mu$\\
    \hline\hline
(I)   &
$s^{-4/3}$ &
$N^{3/4}$ &
$\displaystyle{\frac{1}{2}}$ &
$2$\\ \cline{2-5}
(IIa) &
\multirow{2}{5cm}{$\begin{cases}
s^{-(\eta+\lambda-2)/(\lambda-2)} & (s\ll s_c),\\
k_m^{(\lambda-2)/3 -\eta} s^{-4/3} & (s\gg s_c)
\end{cases}$} &
$N^{(4\lambda-5-3\eta)/[4(\lambda-1)]}$ &
\multirow{2}{1.5cm}{$\displaystyle{\frac{\eta}{\lambda-2-\eta}}$} &
$\displaystyle{\frac{\eta}{1+3\eta} \frac{4(\lambda-1)}{\lambda-2-\eta}}$
 \\
   \cline{3-3} \cline{5-5}
(IIb)&&
$N^{(\lambda-2)/(\lambda-2+\eta)}$ &&
$\displaystyle{\frac{\lambda-2+\eta}{\lambda-2-\eta}}$\\
  \cline{2-5}
(IIIa1) &
\multirow{4}{7cm}{
$\begin{cases}
k_m^{(\lambda-1)(\lambda-2-\eta)/\eta}
s^{-(\lambda-1+\eta)/\eta}& (s\ll s_*)\\
k_m^{(\lambda-2-\eta)(\lambda-2+\eta)/\eta}
s^{-(\lambda-2+\eta)/\eta}& (s_*\ll s\ll s_c)\\
k_m^{(\lambda-2)/3 - \eta} s^{-4/3} & (s\gg s_c).
\end{cases}$}&
\multirow{2}{4cm}{$
N^{(4\lambda-5-3\eta)/[4(\lambda-1)]}$}&
\multirow{4}{1.5cm}{$\displaystyle{\frac{\eta}{\eta-\lambda+2}}$} &
$\displaystyle{\frac{\eta}{1+3\eta} \frac{4(\lambda-1)}{\eta-\lambda+2}}$ \\
    \cline{5-5}
(IIIa2) &&&&
\multirow{3}{3cm}{$\displaystyle{\frac{2 \eta}{\eta-\lambda+2}}$}\\
\cline{3-3}
(IIIb)&&
$N^{\eta/(\lambda-2+\eta)+(\lambda-2-\eta)/(\lambda-1)}$&&\\
\cline{3-3}
(IIIc)&&
$N^{(\lambda-2)/(\lambda-1+\eta)}$&&\\
    \cline{2-5}
(IV) &
$\begin{cases}
k_m^{(\lambda-1)(\lambda-2-\eta)/\eta}
s^{-(\lambda-1+\eta)/\eta}& (s\ll s_c)\\
k_m^{(1-2\lambda)/3} s^{-4/3} & (s\gg s_c).
\end{cases}$ &
$N^{(\lambda-2)/(\lambda-1+\eta)}$&
$\displaystyle{\frac{\lambda-1}{\eta-\lambda+2}}$ &
$\displaystyle{\frac{2(\lambda-1)}{\eta-\lambda+2}}$
\\ \hline
\end{tabular}
\end{table*}

In this section, we first investigate the $N$-dependence of the
largest cluster size at $\couplingcoeff_c$. Next, based on this
result, we derive a finite-size scaling form for the order
parameter near $J_c$.

\subsection{The largest cluster size}

The largest cluster size $S$ can be obtain from the
relation~\cite{percolation}, \be \sum_{s>S} p(s) \sim \frac{S}{N}.
\label{eqn:extreme} \ee

{\bf In the case (I),} we use the result of Eq.~(\ref{eqn:ps1}),
and obtain simply that \be S \sim N^{3/4}. \label{eqn:S1} \ee

{\bf In the case (II),} $p(s)$ displays a crossover at $s_c$, and
thus the obtained value of the largest cluster size must satisfy
the self-consistency conditions. For instance, the largest cluster
size obtained by Eq.~(\ref{eqn:extreme}) for $s\ll s_c$ in
Eq.~(\ref{eqn:ps2}) has to be smaller than $s_c$. As a result, the
largest cluster size behaves differently in the two subcases $\eta
< 1$ and $\eta \ge 1$, which we denote (IIa) and (IIb),
respectively. In each subcase, we obtain that \be S\sim
\begin{cases}
N^{(4\lambda-5-3\eta)/[4(\lambda-1)]} & \text{in (IIa),}\\
N^{(\lambda-2)/(\lambda-2+\eta)} & \text{in (IIb)}.
\end{cases}
\ee The largest cluster size $S$ in (IIa) was determined from
$p(s)$ for $s\gg s_c$, and is indeed much larger than $s_c$,
whereas it in (IIb) was done from $p(s)$ for $s\ll s_c$.

{\bf In the case (III),} $p(s)$ exhibits three distinct power-law
behaviors. Thus, this case is divided into three subcases. They
are as follows: $\eta < 1$ (IIIa), $1 \le \eta \le
\sqrt{\lambda^2-3\lambda+3}$ (IIIb), and
$\eta>\sqrt{\lambda^2-3\lambda+3}$ (IIIc). The largest cluster
size in each subcase is given as \be S\sim
\begin{cases}
N^{(4\lambda-5-3\eta)/[4(\lambda-1)]} & \text{in (IIIa),}\\
N^{\eta/(\lambda-2+\eta)+ (\lambda-2-\eta)/(\lambda-1)} & \text{in (IIIb),} \\
N^{(\lambda-2)/(\lambda-1+\eta)} & \text{in (IIIc).}
\end{cases}
\ee

{\bf In the case (IV),} the largest cluster size is determined
simply by $p(s)$ for $s\ll s_c$ since the resulting largest
cluster size fulfils the criterion $S < s_c$ for $\eta>0$. Thus,
\be S\sim N^{(\lambda-2)/(\lambda-1+\eta)}. \ee

\subsection{Finite-size scaling}
Here, we evaluate the magnitude of the order parameter
$\orderparameter_c$ at $J_c$ and establish the finite-size scaling
function. To proceed, we compare the magnitude of cooperative
synchrony $S/N$ with that of the background synchrony $\sim
N^{-1/2}$. The order parameter $r_c$ is defined as $\sim S/N$ if
$S/N\gg N^{-1/2}$, and $\sim N^{-1/2}$ otherwise. Under this
criterion, we obtain $r_c$ as $\sim S/N$ in the cases (I) and
(II), and $\sim N^{-1/2}$ in the cases (IIIb), (IIIc), and (IV).
The case (IIIa) is divided into two subcases, $2\lambda-3\eta-3
\ge ( < ) 0$. They are denoted as (IIIa1) and (IIIa2),
respectively. The order parameter $r_c$ behaves as $\sim S/N$ and
$\sim N^{-1/2}$ in (IIIa1) and (IIIa2), respectively.

By using that $r\sim \Delta^{\beta}$ and $N$-dependent behavior of
$r_c$ at $J_c$, we can construct a finite-size scaling form as \be
\orderparameter= N^{-\beta/\mu} \psi (
    \Delta N^{1/\mu})
\label{eqn:fss12} \ee for the cases (I) and (II), and \be
\orderparameter= N^{-\beta/\mu} \psi (
    \couplingcoeff N^{1/\mu})
\label{eqn:fss34} \ee for the cases (III) and (IV), where \be
\psi(x)\sim
\begin{cases}
\text{const} & \text{for}~~x\ll 1, \\
x^\beta & \text{for}~~x\gg 1.
\end{cases}
\ee The critical exponent $\mu$ is determined by the relation
$r_c\sim N^{-\beta/\mu}$. The value of $\mu$ varies depending on
the cases determined by the magnitude of $\eta$ and $\lambda$.

We present the diagram in Fig.~\ref{fig:regions} comprising eight
distinct cases in the $(\eta,\lambda)$ plane. Each case in the
diagram corresponds to a distinct behavior of the critical
exponents $\beta$ and $\mu$, the cluster-size distribution, and
the largest cluster size. We summarize those features in
Table~\ref{table:exponents}.

\begin{table*}[t]
\caption{Numerical values of the parameters ($\eta$,\,$\lambda$)
we used for Figs.~\ref{fig:numericalresults}. $\beta_t$ and
$\mu_t$ are theoretical values for a given set of
($\eta$,\,$\lambda$) in the third column. $\beta_n$ and $\mu_n$
are numerical values to draw Figs~\ref{fig:numericalresults} for
each case. For (a)--(c), the theoretical and numerical values are
the same each other for both $\beta$ and $\mu$. However, they can
be different for (d)--(h). $r_t$ and $r_n$ are the order
parameters in scaling form formulated with the theoretical values
of $\beta_t$ and $\mu_t$, and the numerical values $\beta_n$ and
$\mu_n$, respectively.} \label{table:lambdanu}
\begin{tabular}{c|c|c|c|c|c|c|l|l}
\hline
Fig.~\ref{fig:numericalresults} & domain  & ($\eta$,\,\,$\lambda$)& $\beta_t$ & $\beta_n$ & $\mu_t$ & $\mu_n$ &~~~~~~~~~~~~~$\orderparameter_t$ & ~~~~~~~~~~~~~$\orderparameter_n$ \\
\hline\hline
(a) & (I)     & $\left(1/3,4\right)$ & $1/2$  & $1/2$   & $2$      & $2$      & ~~$N^{-1/4}\psi(\Delta N^{1/2})$ & ~~$N^{-1/4}\psi(\Delta N^{1/2})$ \\
(b) & (IIa)   & $\left(5/6,4\right)$ & $5/7$  & $5/7$   &
~$49/120$~ &
~$49/120$~ & ~~$N^{-7/24} \psi(\Delta N^{49/120})$ &~~$N^{-7/24} \psi(\Delta N^{49/120})$ \\
(c) & (IIb)   & $\left(4/3,4\right)$ & $2$    & $2$     & $5$      & $5$      & ~~$N^{-2/5}\psi(\Delta N^{1/5})$ & ~~$N^{-2/5}\psi(\Delta N^{1/5})$ \\
(d) & (IIIa1) & $~\left(1/3,13/6 \right)$ & $2$ & $20/7$  & $14/3$   & $20/3$   & ~~$N^{-3/7}\psi(\couplingcoeff N^{3/14})$ & ~~$N^{-3/7}\psi(\couplingcoeff N^{3/20})$ \\
(e) & (IIIa2)~& $~\left({2}/{3},{13}/{6}\right)$~& ~$4/3$~  & ~$50/23$~ & $8/3$    & $100/23$ & ~~$N^{-1/2}\psi(\couplingcoeff N^{3/8})$ & ~~$N^{-1/2}  \psi(\couplingcoeff N^{23/100})$ \\
(f) & (IIIb)  & $\left({5}/{2},4 \right)$ & $5$    & $5$     & $10$     & $10$     & ~~$N^{-1/2}\psi(\couplingcoeff N^{1/10})$ & ~~$N^{-1/2}\psi(\couplingcoeff N^{1/10})$ \\
(g) & (IIIc)  & $\left({11}/{4},4 \right)$ & $11/3$ & $11/3$  & $22/3$   & $22/3$   & ~~$N^{-1/2}\psi(\couplingcoeff N^{3/22})$ & ~~$N^{-1/2}\psi(\couplingcoeff N^{3/22})$ \\
(h) & (IV)    & $\left(4,4\right)$ & $3/2$  & $5$     & $3$      & $10$     & ~~$N^{-1/2}\psi(\couplingcoeff N^{1/3})$ & ~~$N^{-1/2}\psi(\couplingcoeff N^{1/10})$ \\
  \hline
\end{tabular}
\end{table*}

\begin{figure*}
\resizebox*{!}{8cm}{\includegraphics{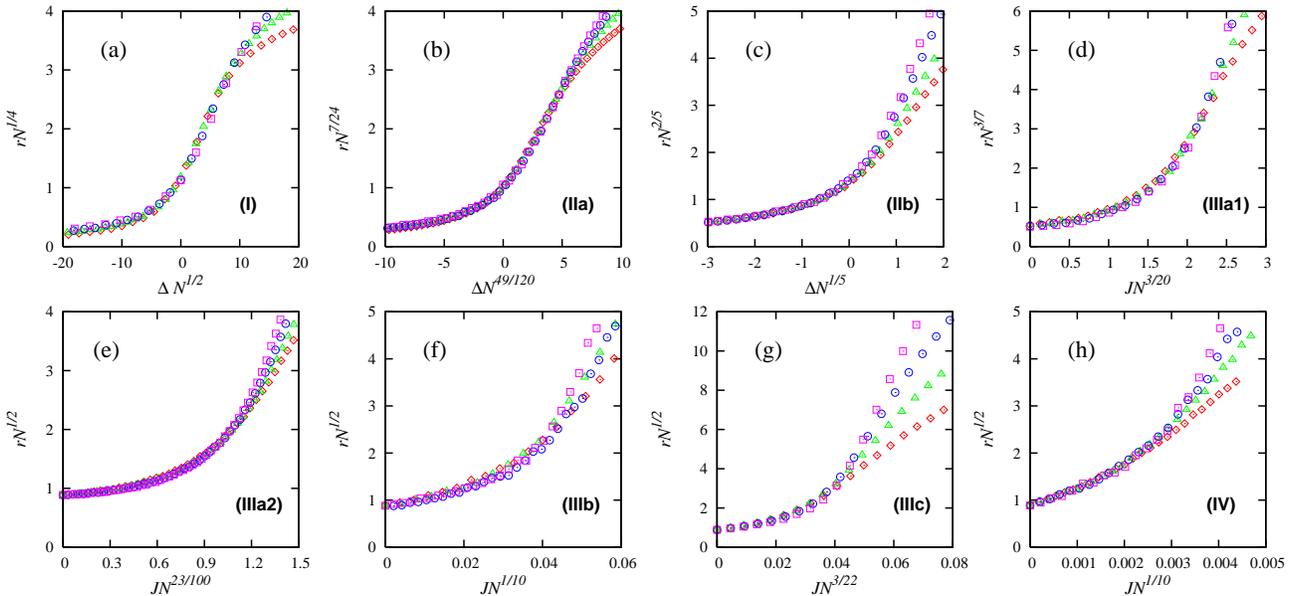}} \caption {(Color
online) Finite-size scaling behaviors of the order parameter $r$.
Data are collected from the static network model with mean degree
$\langle k \rangle = 4$ and system sizes
$N=$400(\textcolor{red}{{\large{$\diamond$}}}),
800(\textcolor{green}{$\triangle$})
1600(\textcolor{blue}{$\bigcirc$}), and
3200(\textcolor{magenta}{$\square$}). Numerical values of the
tunable parameters ($\eta$,\,$\lambda$) and the critical exponents
are given in Table.~\ref{table:lambdanu} for each case. For the
critical point $J_c$, theoretical $J_c^{(t)}$ and numerical
$J_c^{(n)}$ values used in (a)--(c) are different as $(J_c^{(t)},
J_c^{(n)})=(0.92,1.32)$ (a), (0.37,0.50) (b), and (0.13,0.18)
(c).} \label{fig:numericalresults}
\end{figure*}

\section{Numerical simulations}
\label{sec:nu}

We perform direct numerical integration of Eq.~(\ref{eqn:eom}) to
confirm the analytic results. In particular, the finite-size
scaling behaviors in Eqs.~(\ref{eqn:fss12}) and (\ref{eqn:fss34})
are compared. For the purpose, we generate random SF networks
using the static model~\cite{staticmodel} with system sizes of
$N=400, 800, 1600$, and $3200$, mean degree of $\langle k \rangle
= 4.0$ and values of $\eta$ and $\lambda$ chosen from each domain
in Fig.~\ref{fig:regions}. Numerical values of $(\lambda,\eta)$ we
used are listed in Table~\ref{table:lambdanu}. For the numerical
integration, we apply the Heun's method~\cite{heun}. Time is
discretized by a unit step $\delta t=0.005$ and runs up to $t=1.2
\times 10^4$. Ensemble average is taken over ${\cal O}(10^2)\sim
{\cal O}(10^3)$ different configurations of natural frequencies
and network realizations, respectively.

Numerical results are presented in
Fig.~\ref{fig:numericalresults}. For each
Fig.~\ref{fig:numericalresults}(a)--(c), the critical point $J_c$
is finite. We find $J_c$ numerically to make the obtained
numerical data collapsed for different system size $N$ in the
scaling plot with theoretical values of $\beta$ and $\mu$.
Theoretical and numerically found values of $J_c$, denoted as
$J_c^{(t)}$ and $J_c^{(n)}$, respectively, are compared as
$(J_c^{(t)}, J_c^{(n)})=(0.92,1.32)$ for (a), (0.37,0.50) for (b),
and (0.13,0.18) for (c). They belong to the cases (I), (IIa), and
(IIb), respectively.

For Figs.3~(d)--(h), the critical point $J_c$ is zero. For the
cases of (d),(e), and (h), we find that numerical data do not
collapse well in the scaling plot of $rN^{\beta/\mu}$ versus $J
N^{1/\mu}$ with theoretical values of $\beta_t$ and $\mu_t$
tabulated in Table~I. Instead, we adjust numerical values of
$\beta_n$ and $\mu_n$ values empirically to make the obtained
numerical data collapsed. Those empirical values of $\beta_n$ and
$\mu_n$ are compared with the theoretical values as listed in
Table II. The discrepancy in (d) and (e) originates from the
presence of intrinsic degree-degree correlation in the static
model when the degree exponent $2 < \lambda < 3$, while the
theoretical values were obtained under the assumption that the
degree-degree correlation is absent.

\section{Conclusions and Discussion}
\label{sec:conclusion}

In this paper, we have investigated the nature of the
synchronization transition generated by $N$ limit-cycle oscillators
located at random SF networks with degree exponent $\lambda$. The
dynamics is given by a modified Kuramoto equation with the
asymmetric and degree-dependent weighted coupling strength in the
form of $J k_i^{1-\eta}$, where $k_i$ is the degree of vertex $i$.
Depending on the sign and magnitude of $\eta$, the influence of the
hub vertices on the dynamics can be moderated or amplified, and
determines the nature of the synchronization transition. Applying
the mean-field approach to the modified Kuramoto equation, we
derived the critical point, the size distribution of synchronized
clusters, and the largest cluster size at the critical point. The
critical exponents associated with the order parameter and the
finite-size scaling are determined in terms of the two tunable
parameters, ($\eta, \lambda$). All results are summarized in
Table.~\ref{table:exponents}. The parameter space of
($\eta,\lambda$) is divided into eight different domains, in each of
which the transition nature is distinct.

It would be interesting to notice that the critical exponents
$\beta$ and $\mu$ associated with the order parameter and the
finite-size scaling of the synchronization transition depend on
the parameter $\eta$ defined in the coupling strength. The result
is unusual from the perspective of the universality in the
critical phenomena in regular lattice where the details of the
couplings are mostly irrelevant unless they are long
ranged~\cite{fisher}. This result implies that structural features
of SF networks such as the degree distribution are not sufficient
to understand the dynamic process on such networks. Asymmetric
coupling in dynamics is a relevant perturbation in such networks,
because of the heterogeneity of the degree distribution. Such a
behavior was also observed in the sandpile model~\cite{sandpile}.

\section{Acknowledgement}
This work was supported by KRF Grant No.~R14-2002-059-010000-0 of
the ABRL program funded by the Korean government (MOEHRD) and the
Seoul R \& BD program.
\appendix

\section{Derivation of Eqs.~(\ref{eqn:localrs}-\ref{eqn:globalrs})}
\label{app:singularity}

In this Appendix, we present the derivation of
Eqs.~(\ref{eqn:localrs}--\ref{eqn:globalrs}) from
Eqs.~(\ref{eqn:localfield-expand}) and
(\ref{eqn:orderparameter-expand}) with $\eta>0$. Since the
right-hand-side of Eq.~(\ref{eqn:orderparameter-expand}) becomes
the same as that of Eq.~(\ref{eqn:localfield-expand}) when
$\lambda$ is shifted by 1, we here analyze in detail the nature of
Eq.~(\ref{eqn:localfield-expand}) only, which applies also to
Eq.~(\ref{eqn:orderparameter-expand}) with $\lambda$ replaced by
$\lambda+1$.

The coefficients in Eq.~(\ref{eqn:localfield-expand})
diverge with increasing $N$ for $n\ge n_c\equiv\lceil
(\lambda-\eta-2)/(2\eta)\rceil$, where $\lceil x\rceil$
is the smallest integer not smaller than $x$, due to
the divergence  of the generalized harmonic
number $H_m^q$ for $0<q<1$. While $H_m^q \simeq \zeta(q)$ for $q>1$,
$H_m^q$ diverges in the limit $m\to\infty$ for $0<q<1$ and
its asymptotic expansion can be obtained by using
the relation $H_m^q = \zeta(q)-\zeta(q,m+1)$
and the asymptotic expansion of the Hurwitz zeta function~\cite{wolfram}
\begin{align}
\zeta(q,m) &= \frac{1}{q-1} m^{1-q}
    +\frac{1}{2}m^{-q}\nonumber\\
&+ 2 m^{1-q}
    \int_0^\infty dx \frac{\sin (q \tan^{-1}x)}
{(1+x^2)^{q/2} (e^{2\pi m x}-1)}.
\label{eqn:hurwitz}
\end{align}
One can see that the integral is of order $m^{-1}$
in the limit $m\to\infty$
since
\begin{align}
&\int_0^\infty dx \frac{\sin (q \tan^{-1}x)}
{(1+x^2)^{q/2} (e^{2\pi m x}-1)} \nonumber\\
&~~\lesssim \int_0^{1/m} dx \frac{q x}{e^{2\pi  m x}-1} +
\int_{1/m}^\infty dx \frac{1}{e^{2\pi  m x}-1}
 \nonumber\\
&~~\simeq {\cal O}(m^{-2}) + {\cal O}(m^{-1}),
\end{align}
and thus for $0<q<1$, $H_m^q$ behaves as
\be
H_m^q = \sum_{k=1}^{m} \frac{1}{k^q}  \simeq
\frac{(m+1)^{1-q}}{1-q} + \zeta(q) +{\cal O}(m^{-q}).
\ee

When $\eta>0$, the terms with such diverging coefficients exist
and thus we can rearrange the expansion as follows:
\begin{align}
\label{eqn:rpositive}
\localfield &=
\sum_{n=0}^{\infty}
      \frac{ (n-1/2)! (-1)^n (\couplingcoeff \localfield)^{2n+1}
      H_{k_m}^{\lambda-\eta-2n\eta-1}}
{2^{n+3/2}n!(n+1)! \zeta(\lambda-1)}
        \\ \nonumber
  &= \sum_{n=0}^{\infty}
      \frac{ (n-1/2)! (-1)^n \zeta(\lambda-\eta-2\eta n -1)}{
    2^{n+3/2} n! (n+1)! \zeta(\lambda-1)}
    (\couplingcoeff \localfield)^{2n+1}\\ \nonumber
  &~~+ \sum_{n=n_c}^{\infty}
      \frac{ (n-1/2)!(-1)^n k_m^{2+\eta +2n \eta-\lambda}
     (\couplingcoeff \localfield)^{2n+1}  }
{ 2^{n+3/2} n! (n+1)!(2+\eta+2n \eta-\lambda) \zeta(\lambda-1)}
\nonumber\\
&=\sum_{n=0}^\infty \bar{B}_n (\couplingcoeff \localfield)^{2n+1}
+ (\couplingcoeff \localfield)^{(\lambda-2)/\eta}
\bar{C}(\couplingcoeff \localfield k_m^\eta),
\end{align}
where we approximated $H_{k_m}^{\lambda-1}$ by $\zeta(\lambda-1)$
since $\lambda-1>1$. The coefficients $\bar{B}_n$ are defined in
Eq.~(\ref{eqn:BBbar}) and the function $\bar{C}(x)$ is defined by
\begin{align}
\bar{C}(x)&= \sum_{n=n_c}^\infty \bar{c}_n
x^{(2+\eta+2n \eta -\lambda)/\eta}, \text{~~~~~~with} \nonumber\\
\bar{c}_n &=
      \frac{ (n-1/2)!(-1)^n }
{  2^{n+3/2} n! (n+1)!(2+\eta+2n \eta-\lambda) \zeta(\lambda-1)}.
\end{align}
While $\bar{C}(x)$ behaves as $x^{(2+\eta-\lambda+2\eta
n_c)/\eta}$ for $x\ll 1$, it converges to a constant
$\bar{C}_\infty$ for $x\to\infty$ yielding a non-analytic term
$\bar{C}_\infty (\couplingcoeff \localfield)^{(\lambda-2)/\eta}$
as can be seen in Eq.~(\ref{eqn:local-pos-eta}). Therefore, the
magnitude of $\couplingcoeff \localfield k_m^\eta$ is essential
for the determination of the leading behaviors of the
right-hand-sides of Eqs.~\sceqs  for $\localfield\ll 1$. If
$\couplingcoeff \localfield k_m^\eta\gg 1$,
  one can approximate
the function $\bar{C}(\couplingcoeff \localfield k_m^\eta)$ by
a constant $\bar{C}_\infty$ that is evaluated as
\begin{align}
\bar{C}_\infty &=\lim_{x\to\infty}
\sum_{n=n_c}^{\infty}
      \frac{ (n-1/2)!(-1)^n 2^{-(\lambda+2\eta-2)/(2\eta)}
}
{n! (n+1)!(2+\eta+2n \eta-\lambda) \zeta(\lambda-1)}
    \nonumber\\
&~~\times (2^{-1/2} x)^{(2+\eta+2n \eta -\lambda)/\eta}\nonumber\\
&=
   \frac{1}{2^{(\lambda+2\eta-2)/(2\eta)}
     \zeta(\lambda-1)}\sum_{n=n_c}^{\infty} \frac{ (-1)^n (n-1/2)! }{n! (n+1)!}
\nonumber\\
    &~~ \times
    \int_{0}^{\infty} dy y^{1+\eta+2n\eta-\lambda}  \nonumber\\
&=
\displaystyle{
\frac{[(\lambda-2\eta-2)/2\eta]![(2-\lambda-\eta)/2\eta]!} {\eta
2^{(\lambda+4\eta-2)/2\eta}[(\lambda+\eta-2)/2\eta]! \zeta(\lambda-1)}}.
\end{align}
The function $C(x)$ defined in the text can be approximated in the
same way by a constant $C_\infty$ that is identical to
$\bar{C}_\infty$ with $\lambda+1$ in place of $\lambda$.
Therefore, one should refer to Eqs.~(\ref{eqn:local-pos-eta}) and
(\ref{eqn:orderparameter-pos-eta}) for the correct expansions of
$\orderparameter$ and $\localfield$ around $\localfield=0$ in the
case of $\couplingcoeff \localfield k_m^\eta\gg 1$ while
Eqs.~\sceqs can be used in the case of $\couplingcoeff \localfield
k_m^\eta\ll 1$.

\end{document}